\documentstyle[aps,prl,multicol,amssymb,epsf]{revtex}

\def\mathbold{\bf}
\def\be{\begin{equation}}
\def\ee{\end{equation}}
\def\bea{\begin{eqnarray}}
\def\eea{\end{eqnarray}}
\def\br{{\mathbold r}}

\newcommand{\corr}[1]{\langle #1\rangle}
\newcommand{\Tr}{\mathop{\rm Tr}}
\newcommand{\pp}{{\textstyle\frac\pi2}}
\newcommand{\ETh}{E_{\rm Th}}

\begin{document}

\draft

\title{Proximity Action theory of superconductive nanostructures}
\author{M. A. Skvortsov$^1$, A. I. Larkin$^{1,2}$ and M. V.
Feigel'man$^1$}

\address{
$^1$L. D. Landau Institute for Theoretical Physics, Moscow
117940, Russia\\
$^2$Theoretical Physics Institute, University of Minnesota,
Minneapolis, MN 55455, USA
}

\maketitle

\begin{abstract}
We review a novel approach to the superconductive proximity effect
in disordered normal--super\-conducting (N-S) structures.
The method is based on the multicharge Keldysh action and is suitable
for the treatment of interaction and fluctuation effects.
As an application of the formalism, we study the subgap conductance
and noise in two-dimensional N-S systems in the presence
of the electron-electron interaction in the Cooper channel.
It is shown that singular nature of the interaction correction
at large scales leads to a nonmonotonuos temperature, voltage
and magnetic field dependence of the Andreev conductance.
\end{abstract}


\begin{multicols}{2}

\section{Introduction}

A superconductor in contact with a normal metal induces Cooper
correlations
between electrons in the normal region, the phenomenon known as the
proximity
effect. Its microscopic origin lies in Andreev reflection~\cite{Andreev}
of
an electron into a hole at the normal-metal--superconducting (N-S)
interface.
The probability of Andreev reflection and thus the strength of the
proximity
effect is determined by the transparency of the N-S interface and the
nature
of electron propagation in the N part of the structure.
Disorder in the normal conductor near the N-S contact was shown
theoretically~\cite{volkov,Nazarov94T,Nazarov94C,Carlo}
to increase considerably the effective probability of Andreev reflection
(see Ref.~\cite{Panne} for a recent review from the experimental
viewpoint).

The standard semiclassical theory of N-S
conductivity~\cite{volkov,Nazarov94T,Nazarov94C,Carlo}, 
based either on the traditional nonequilibrium 
superconductivity approach~\cite{LOreview}
or on the scattering formalism~\cite{scattering,Carlo},
usually neglects interaction effects in the N part of the structure.
However, in low-dimensional structures, Coulomb interaction in the
normal diffusive region gets enhanced~\cite{AA}, which may affect strongly
the Andreev conductance and noise.

In this paper we address the effect of interaction between electrons in
the
normal part of an N-S structure on the charge transport through the
system.
To study a system with interaction a novel theoretical method should be
developed since neither of the above-mentioned approaches can handle
interaction corrections. Indeed, the scattering matrix formalism
relying on the linear relation between the outgoing and incoming
states
is {\em a priori} a one-particle description.
On the other hand,
Larkin-Ovchinnikov kinetic equation~\cite{LOreview} can be generalized
to allow for (at least some part of) interaction corrections, but
its practical solution seems hardly possible beyond the first order of
perturbation theory in interaction strength~\cite{Spivak,StNaz}.

\begin{figure}
\refstepcounter{figure} \label{F:NSisland}
\vspace{1mm}
\epsfxsize=3in
\centerline{\epsfbox{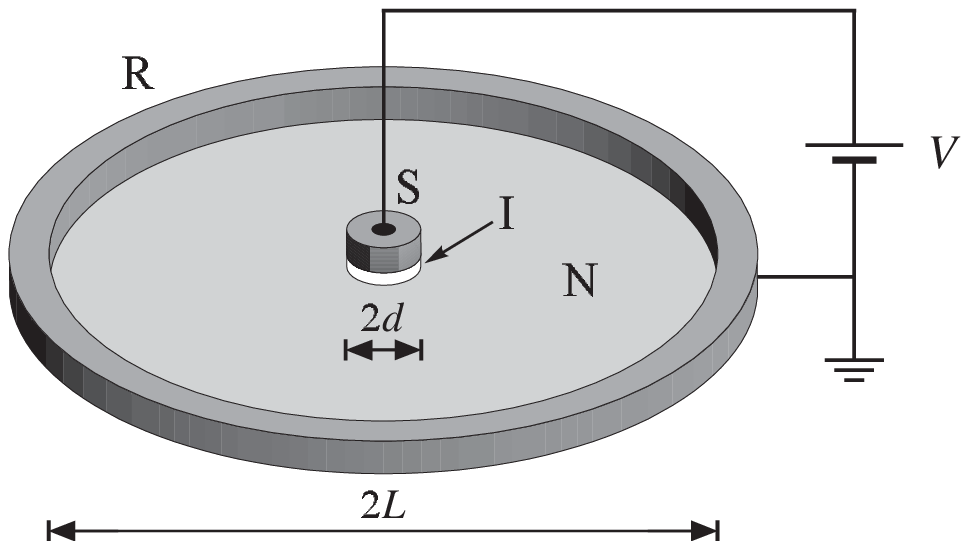}}
\vspace{3mm}
\small FIG.\ \arabic{figure}.
A small superconductive island (S) of size $2d$ connected to
a reservoir (R) through a tunnel barrier (I) and
a dirty normal film (N) of size $2L \gg 2d$.
\end{figure}

An appropriate formalism had been developed in Ref.~\cite{SLF} in the
framework of the Keldysh action for disordered
superonductors~\cite{FLS}.
We start from the fully microscopic Lagrangian describing interacting
electrons in the diffusive conductor. Then, successively integrating
over electronic degrees of freedom in the normal conductor we end up
with the {\em Proximity Action}, $S_{\rm prox}[Q_S,Q_N]$, which is
a functional of two matrices, $Q_S$ and $Q_N$, describing the states
of the superconductive and external normal terminals of the N-S
structure
(cf.\ Fig.~\ref{F:NSisland}).
Once the form of the Proximity Action is known, one can easily calculate
the conductivity of the system, current noise, and, in principle,
higher correlators of current and even the full statistics of
transmitted
charge~\cite{LeviLes,LLYa}.

The Proximity Action approach bears an obvious analogy
with the scattering matrix approach~\cite{Carlo,Nazarov94T}
as both describe transport properties in terms of the characteristics
of the terminals (stationary-state Green functions of the terminals
$Q_{S,N}$
in the former versus asymptotic scattering states in the latter
approach).
In this respect, the Proximity Action method also shares the logic
of Nazarov's circuit theory of Andreev conductance~\cite{Nazarov94C}.
On the other hand, the Keldysh action approach is a natural
generalization
of the kinetic equation for dirty superconductors for the case
of fluctuating fields. The Larkin-Ovchinnikov kinetic equation then
emerges as a saddle point equation for the Keldysh action~\cite{FLS}.

As an application of the formalism, we will study charge transport
in two-dimensional (2D) N-I-S structures shown in Fig.~\ref{F:NSisland}
at low (compared to the S gap $\Delta$) temperature and voltages,
and arbitrary ratio $t=R_D/R_T$, where $R_D$ and $R_T$ are the
resistances
of the diffusive normal conductor, and of the tunnel barrier
in the normal state, correspondingly.
We will calculate the Andreev conductance and noise of such systems
in the presence of Cooper interaction in the normal conductor
modified by the Coulomb interaction~\cite{finkel1,finkel2,finkel3},
as a function of the ``decoherence time"
of an electron and the Andreev-reflected hole, $\hbar/\Omega_*$,
where $\Omega_* = \max(T, eV, eDH/c)$.
We will show that the Cooper interaction effects in 2D are substantially
different from those in the 1D case considered in Ref.~\cite{StNaz}.
In particular, the lowest-order relative correction $\delta G_A/G_A$
scales as $-\lambda\ln(L/d)$ and grows with the size of the system $L$,
provided that $\Omega_* \ll \ETh = \hbar D/L^2$.
We will sum up the main logarithmic terms of the order of
$[\lambda\ln(L/d)]^n$ and find that Cooper repulsion may lead
to a nonmonotonuos dependence of $G_A$ on $R_D/R_T$ and on
the decoherence energy scale $\Omega_*$.

In this paper we will not consider the weak
localization~\cite{band4,GLK}
and interaction-induced corrections~\cite{AA} to the sheet conductance,
$\sigma$, of the normal metal assuming that it is relatively large,
$g\equiv(\hbar/e^2)\sigma \gg \ln(L/d)$.
We will also not take into account the effect of the Coulomb
zero-bias anomaly (ZBA)~\cite{AA,AAL} on the Andreev conductance.
This effect was considered in~\cite{FLS} within the lowest-order
tunneling approximation. It was shown, similar to the Coulomb ZBA
problem for normal conductors (cf., e.~g., Ref.~\cite{LS}), 
that it can be strongly suppressed if the bare
long-range Coulomb potential is screened, e.~g.,
by placing a nearby metal gate. This is the situation we assume
in the present paper.

Another limitation of the present discussion is that we will consider
the case of a 2D geometry of the current flow
between the superconductive and normal electrodes of the structure, as
shown in Fig.~\ref{F:NSisland}. This will make possible to construct
a unified functional renormalization group treatment that takes into
account modifications of the proximity effect strength both due to
multiple Andreev reflections and due to Cooper-channel repulsion.
Note that the sample geometry should not necessarily be symmetric as
shown in Fig.~\ref{F:NSisland}: the only
important feature of the geometry chosen is the 2D spreading of
the current flow.

\section{Multicharge Proximity Action}

Here we present a brief overview of the Proximity Action approach
to charge transport in N-I-S systems elaborated in Ref.~\cite{SLF}.
Eliminating all dynamic degrees of freedom in the normal diffusive
conductor, one reduces the total action of the system to the form
$S_{\rm prox}[Q_S,Q_N]$, with the matrices $Q_S$ and $Q_N$ having
the meaning of the steady-state Keldysh Green function
of the S island and external normal terminal labeled by R
in Fig.~\ref{F:NSisland}, respectively.
In the low frequency limit, $\omega \ll \ETh = D/L^2$, when diffusive
electron motion spreads over the whole conducting region during
one circle of the external voltage, the resulting Proximity Action
has the form
\be
  S_{\rm prox}[Q_S,Q_N]
  = -i \pi^2 g \sum_{n=1}^\infty \gamma_n \Tr (Q_S Q_N)^n .
\label{prox}
\ee
The action is known once the infinite set of parameters (``charges'')
$\gamma_n$ is specified.

Remarkably, for noninteracting systems, knowledge of $\{\gamma_n\}$
is equivalent to knowledge of the whole distribution function
${\cal P}(T)$ of transmission coefficients of the system.
To formulate this analogy, it is convenient to make a Fourier transform
from $\{\gamma_n\}$ to a $2\pi$-periodic function $u(x)$ of an
auxiliary continuous variable $x$:
\be
  u(x) = \sum_{n=1}^\infty n \gamma_n \sin nx .
\label{u-def}
\ee
The relation between the function $u(x)$ and the generating function
of transmission coefficients introduced by Nazarov~\cite{Nazarov94T}
(cf.\ also Ref.~\cite{Carlo94}) was established in Ref.~\cite{SLF} by
direct comparison:
\be
  4\pi g\, u(x)
  = \int_0^1 \frac{T\sin x}{1-T\sin^2(x/2)}\, {\cal P}(T)\, dT.
\label{equiv}
\ee
For interacting systems and systems with fluctuations \cite{FLS-prl},
Eq.~(\ref{equiv}) may be interpreted as a generalization of the notion
of transmission coefficients.

Determination of $\{\gamma_n\}$ is a separate
task related with elimination of fast modes in the normal conductor.
We will discussed it later, while now we will explain how to obtain
transport properties from the Proximity Action (\ref{prox}).
In order to study conducting properties of the system, we
apply a bias voltage $V$ to the superconducting and normal
terminals of the structure and study its current response.
The resulting expressions for the Andreev conductance
(in units of $e^2/\hbar$) and noise power take the form~\cite{SLF}
\be
  G_A = 4\pi g\, u_x (\pp) ,
\label{Gs}
\ee
and
\bea
  \corr{I_{\omega}I_{-\omega}} &=&
  \frac{e^2G_A}{3\hbar} \,
  \Bigl\{
    (3-P_S) \, \Psi(\omega)
\nonumber \\
  {} &+&
    \frac12 P_S \, [\Psi(\omega-2eV) + \Psi(\omega+2eV)]
  \Bigr\} ,
\label{IIs}
\eea
where $\Psi(\omega) = \omega \coth (\omega/2T)$,
and the superconductive noise function $P_S$ is given by
\be
  P_S = 1 - \frac{u_{xxx}(\pp)}{2u_x(\pp)}.
\label{Ps}
\ee
The quantity $F=(2/3)P_S$ gives the reduction of the
shot noise between the normal and superconducting terminals,
$\corr{II}_{\rm shot} = F\, e\corr{I}$,
compared to its Poissonian value for a single electron tunneling,
and can be identified with the Fano factor~\cite{Blanter}.

For an N-S system, all quantities are expressed in terms of derivatives
of $u(x)$ at $x=\pi/2$. The counterparts of Eqs.~(\ref{Gs})--(\ref{Ps})
for the case when the S island is in the normal state
can also be expressed in terms of the same function $u(x)$,
but with derivatives taken at $x=0$.
The conductance of the system $G = 4\pi g\, u_x (0)$,
while the current-current correlator is given by Eq.~(\ref{IIs})
with the replacements $G_A \to G$, $2eV \to eV$,
and $P_S \to P_N = 1 - 2u_{xxx}(0)/u_x(0)$.

In principle, one can go further and calculate the full charge
transfer statistics in terms of the function $u(x)$
following the derivation~\cite{LeviLes,LLYa} in the
scattering matrix technique.

\section{Functional RG}

We have seen that the set $\{\gamma_n\}$ or, equivalently, the function
$u(x)$ encodes all information about charge transmission through the
system.
Now we will discuss how to determine the charges $\gamma_n$ for the
system shown in Fig.~\ref{F:NSisland}.

The starting point is the microscopic Keldysh action for a S island
in contact with a disordered metal, derived in Ref.~\cite{FLS}.
It can be represented as a sum of the bulk and boundary contributions.
The bulk action, $S_{\rm bulk}$, is a functional of three fluctuating
fields: the matter field $Q(\br,t,t')$ in the film,
the electromagnetic potential, and the order parameter field
used to decouple the quartic interaction vertex in the Cooper channel.
$Q(\br,t,t')$ is a matrix in the direct $4\times4$-dimensional
product $K\otimes N$ of the Keldysh and Nambu-Gor'kov spaces,
its average value giving the electron Green function in the
Keldysh form, $\hat{G}({\bf r,r})$.
In the tunneling Hamiltonian approximation, the boundary action,
\be
  S_{\rm tun}[Q_S,Q]= - \frac{i\pi G_T}{4} \Tr Q_S Q(d),
\label{tun}
\ee
describes an elementary tunneling process between the S island and
the N metal. Here
 $G_T=\hbar/e^2R_T$ is the dimensionless tunneling conductance of
the interface, and $Q(d)$ is taken at the metal side of the N-S
boundary.

The next step is to eliminate degrees of freedom in the N film.
2D geometry of the system suggests that this procedure
can be realized with the help of the Renormalization Group (RG)
approach by successive integrating over fast modes in the bulk.
As a result, the boundary action (\ref{tun}) which initially described
single Andreev reflection process gets modified by multiple Andreev
reflections and acquires the form
\be
  S_{\rm bound}
  = -i \pi^2 g \sum_{n=1}^\infty \gamma_n(\zeta) \Tr [Q_SQ(r)]^n .
\label{sg}
\ee
where the logarithmic variable $\zeta=2\ln(r/d)=\ln(\omega_d/\Omega)$,
$\omega_d=D/d^2$ is the high-frequency cutoff, and $\Omega=D/r^2$
is the energy associated with the current scale $r$.
At the energy scale $\omega_d$, the multicharge action (\ref{sg})
reduces to Eq.~(\ref{tun}):
$\gamma_1(0)=a \equiv G_T/4\pi g$, and $\gamma_{n\geq2}(0)=0$,
and, consequently, $u(x,\zeta=0)=a\sin x$.

Under the action of the RG, the charges $\gamma_n(\zeta)$ are changing.
In the zero-energy limit, when the Cooperon decoherence energy
$\Omega_*$ is smaller than the current RG energy scale $\Omega$,
both diffusons and Cooperons contribute to the RG equation
which, written in terms of the function $u(x,\zeta)$, reads~\cite{SLF}
\be
  u_\zeta + u u_x = - \lambda(\zeta)\, u(\pp,\zeta) \sin x.
\label{urg}
\ee
The r.h.s.\ of the functional RG equation~(\ref{urg}) is due to
the Cooper channel interaction with the scale-dependent constant
$\lambda(\zeta)$ given by~\cite{finkel2,FLS}
\be
\label{lambda-res}
  \lambda(\zeta) =
    \frac{\lambda_d+\lambda_g\tanh\lambda_g \zeta}
      {\displaystyle 1+\frac{\lambda_d}{\lambda_g}\tanh\lambda_g \zeta}
,
\ee
where $\lambda_d$ is defined at the energy scale $\omega_d$.
At $\zeta > \sqrt{g}$, Coulomb repulsion in 2D drives $\lambda(\zeta)$
toward the Fikelstein's fixed point $\lambda_g=1/2\pi\sqrt{g}$.
In deriving Eq.~(\ref{urg}) we neglected the weak localization
and interaction corrections to the film conductance $g$,
making use of the small parameter $\zeta/g \ll 1$.

In the absence of Cooper channel interaction, Eq.~(\ref{urg})
acquires the form of the Euler equation.
In this case it is convenient to rewrite it in terms of the new
variable $t = R_D/R_T = a\zeta$
and the function $U(x,t)=u(x,\zeta)/a$
as $U_t+UU_x=0$, with the initial condition $U(x,0)=\sin x$.
In this representation, all information about geometry of the
system had gone, and the Euler equation for $U(x,t)$ describes evolution
of $\{\gamma_n\}$ for any (not only 2D) system consisting of
a tunnel barrier and a diffusive conductor in the zero-energy
limit~\cite{Carlo94,Naz-Ohm,SLF}.

If the decoherence energy scale $\Omega_* > \ETh$ then Cooperons
become inoperative at the latest stage of renormalization procedure,
where $\zeta>\zeta_* \equiv \ln(\omega_d/\Omega_*)$.
In this case the functional RG equation reduces to
\be
  \tilde u_\zeta + \tilde u \tilde u_x = 0,
\label{turg}
\ee
where $\tilde u(x) = [u(x)-u(\pi-x)]/2$.
The function $\tilde u(x,\zeta;\zeta_*)$ bears the knowledge
of the decoherence scale $\Omega_*$ through the initial condition
$\tilde u(x,\zeta_*;\zeta_*) = [u(x,\zeta_*)-u(\pi-x,\zeta_*)]/2$,
where $u(x,\zeta_*)$ is the solution of Eq.~(\ref{urg})
at the upper border of its applicability, $\zeta=\zeta_*$.
Note, however, that crossover from the $\zeta<\zeta_*$ to
$\zeta>\zeta_*$ regimes is described by Eqs.~(\ref{urg}),
(\ref{turg})  with the logarithmic accuracy only.

The RG equations (\ref{urg}) and (\ref{turg}) should be
integrated until we reach the perfect boundary between
the N film and the reservoir.
Then, substituting $Q(L)=Q_N$ into the effective boundary
action (\ref{sg}), we arrive at the Proximity Action
(\ref{prox}) with $\gamma_n=\gamma_n(\zeta_{\rm Th})$, where
$\zeta_{\rm Th} = \ln(\omega_d/\ETh) = 2\ln(L/d)$.

\section{Andreev conductance}

Now we proceed with the analysis of the functional RG equations
derived in the previous section. We start from the zero-energy
limit, $\Omega_* < E_{\rm Th}$, described by Eq.~(\ref{urg}).
In the absence of interaction one readily reproduces the known
result~\cite{Nazarov94T,Nazarov94C} for the Andreev conductance:
\be
  G_A^{(0)}
  = G_T \frac{\sin\Theta(t)}{1+t\sin\Theta(t)}
  = G_D \frac{t\sin\Theta(t)}{1+t\sin\Theta(t)} ,
\label{GA}
\ee
where $t=a\zeta_{\rm Th}=R_D/R_T$, $R_D=\ln(L/d)/2\pi\sigma$,
and the function $\Theta(t)$ satisfies $\Theta(t) = t \cos \Theta(t)$.

\begin{figure}
\refstepcounter{figure} \label{F:R/R1}
\epsfxsize=80mm
\centerline{\epsfbox{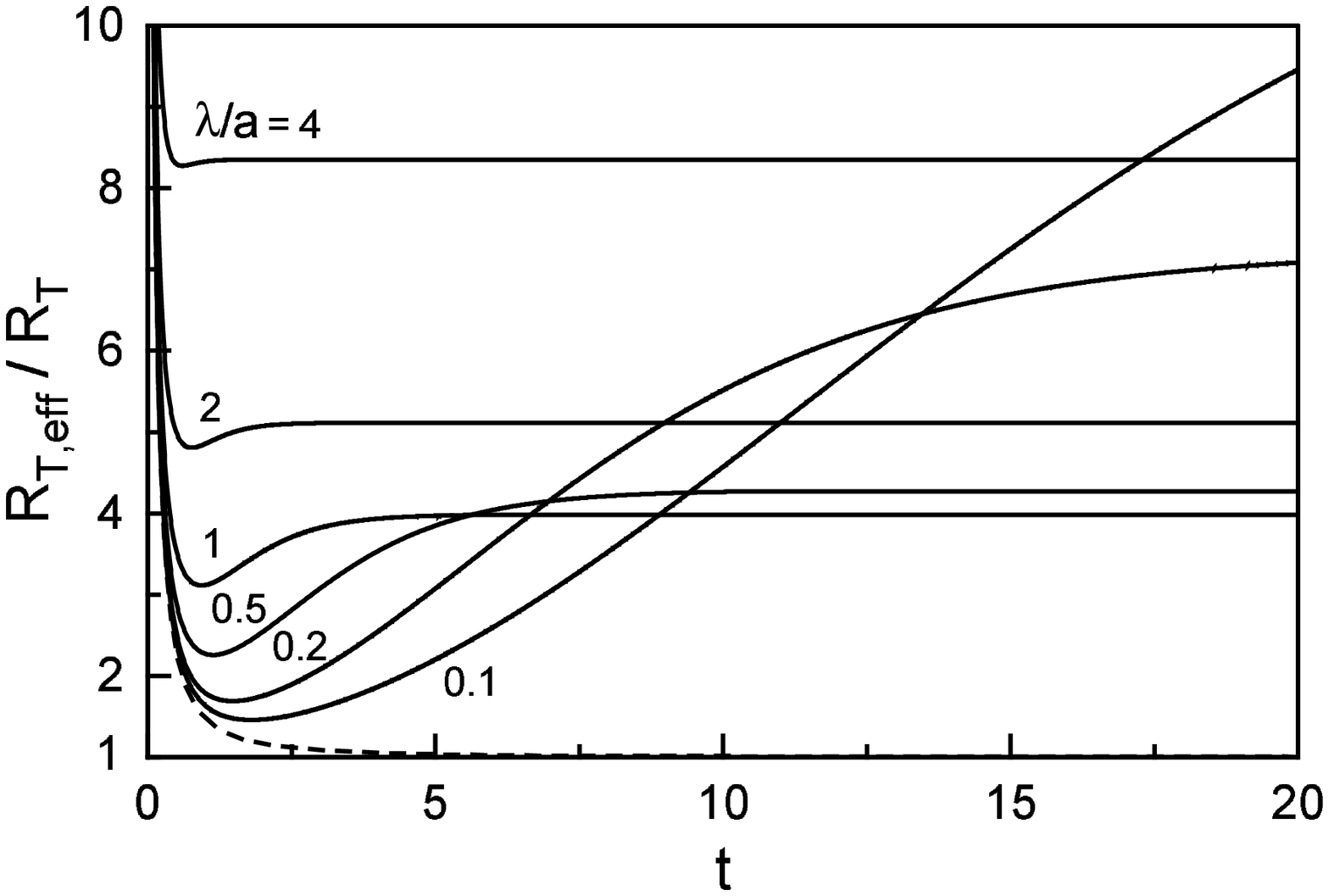}}
\small FIG.\ \arabic{figure}.
Dependence of the effective interface resistance $R_{T,\rm eff}(t)/R_T$
vs.\ $t=R_D/R_T$ for different values of $\lambda/a$ obtained by
numerical
solution of Eq.~(\protect\ref{urg}) for $\lambda(\zeta)=\mbox{const}$.
The dashed line corresponds to the noninteracting case, $\lambda=0$.
\end{figure}

It is often convenient to characterize the system with the help
of the {\em effective interface resistance}\/ $R_{T,\rm eff}$
defined formally through $R_A=R_D+R_{T,\rm eff}$.
For a noninteracting system,
the ratio $R_{T,\rm eff}/R_T$ as a function of $t=R_D/R_T$
is shown in Fig.~\ref{F:R/R1} by the dashed line.
It behaves as $t^{-1}$ for $t\ll1$ and saturates at $1$ for $t\gg1$.
Correspondingly, $G_A^{(0)}=G_T^2/G_D$ for $t\ll1$,
and $G_A^{(0)}=G_D$ for $t\gg1$.

At relatively small scales, the effect of Cooper interaction described
by the r.h.s.\ of Eq.~(\ref{urg}) can be considered perturbatively.
The magnitude of the first-order interaction correction to the
quasiclassical conductance (\ref{GA}),
$\delta G_A \sim -\lambda\zeta G_A^{(0)}$,
{\em grows}~\cite{SLF} with the system size $L$ and eventually becomes
of the order of $G_A^{(0)}$ at $\zeta\sim1/\lambda$.
Comparing this scale to the scale $\zeta\sim 1/a$,
corresponding to the crossover from the tunneling ($t\ll1$)
to diffusive ($t\gg1$) limits, one can distinguish between the regimes
of strong, $\lambda \gg a \equiv G_T/4\pi g$, and weak, $\lambda \ll a$,
repulsion (still, in both cases $\lambda\ll1$).

In general, the functional RG equation (\ref{urg}) does not allow
an analytical solution, and should be solved numerically.
For clarity, we will assume hereafter that the interaction constant
had reached the Finkelstein's fixed point $\lambda_g = 1/2\pi\sqrt{g}$,
so that strong repulsion corresponds to a relatively weak tunneling
conductance $G_T \ll 2\sqrt{g}$, and vice versa.
The effective interface resistance $R_{T,\rm eff}$
normalized to the tunneling resistance $R_T$
as a function of $t=R_D/R_T$ is plotted in Fig.~\ref{F:R/R1}
for different values of the ratio $\lambda_g/a$.
For the case of strong repulsion, $\lambda_g\gg a$,
$R_{T,\rm eff}(t)$ very quickly (at $t\sim a/\lambda_g\ll1$)
reaches its asymptotic value
and saturates at $R_{T,\rm eff}(t=\infty) \approx(2\lambda_g/a)R_T$.
The limiting value $R_{T,\rm eff}(\infty)$ decreases with
the decrease of $\lambda_g/a$ up to $\lambda_g/a \sim 1$.
At smaller $\lambda_g/a$, corresponding to the case of weak repulsion,
$R_{T,\rm eff}(\infty)$ starts to grow again, and
reaches an asymptotic behavior
$R_{T,\rm eff}(\infty)/R_T \approx 1.19a/\lambda_g$
at $\lambda_g/a \ll 1$.
In this limit $R_{T,\rm eff}(t)$ reaches its asymptotic value
at large scale $t\sim a/\lambda_g$.

The data of Fig.~\ref{F:R/R1} demonstrate that in a 2D system
the limits $\lambda\to0$ and $R_D\to\infty$ {\em do not commute}:
for any small but finite $\lambda_g$,
$R_{T,\rm eff}(t)/R_T$ will eventually (though, at very large $t$)
deviate from the noninteracting dependence
(dashed line in Fig.~\ref{F:R/R1}) and become large.

\begin{figure}
\refstepcounter{figure} \label{F:ga_t_int}
\epsfxsize=80mm
\centerline{\epsfbox{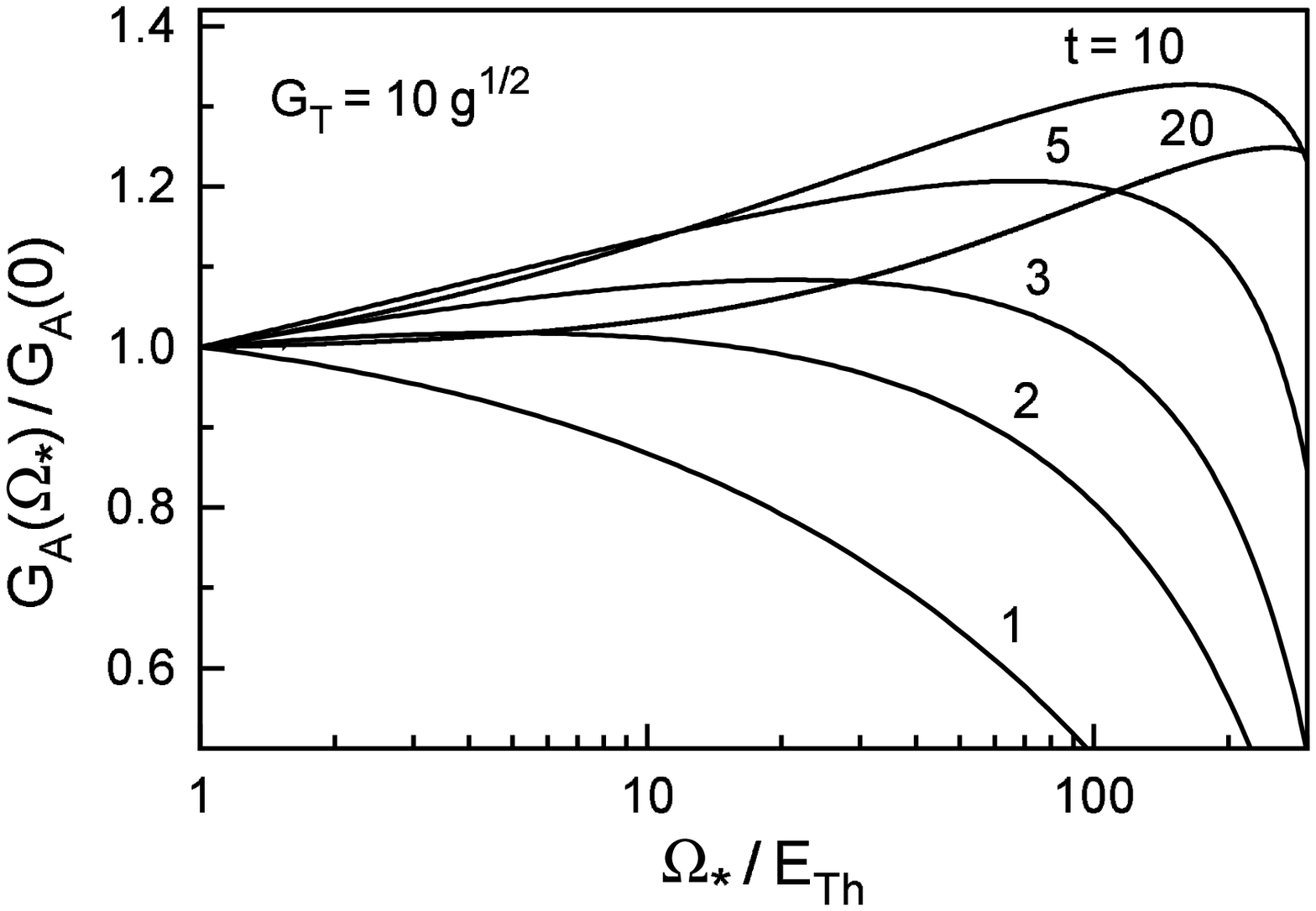}}
\small FIG.\ \arabic{figure}.
Dependence of the Andreev conductance $G_A(\Omega_*)$ (normalized to
the zero-energy value) on the ratio $\Omega_*/\ETh$
for different values of $t$. For all plots, $\omega_d/\ETh=400$,
$\lambda=\lambda_g$, and $G_T=10\sqrt{g}$.
\end{figure}

At large enough temperature, voltage, or magnetic field,
$\Omega_*\equiv\max(T, eV, eDH/c) \gg \ETh$, the resistance
of the structure is given by~\cite{SLF}
\be
  R_A(t,t_*) = R_D + R_{T,\rm eff}(t_*),
\label{RA*}
\ee
where $t_*=a\zeta_*=(G_T/4\pi g)\ln(\omega_d/\Omega_*)$.
Therefore, a nonmonotonous behavior of $R_{T,\rm eff}(t_*)$
will manifest itself in a nonmonotonous temperature, voltage
and magnetic field dependence of the subgap conductance.
The unusual enhancement of conductivity with
the increase of the decoherence energy scale $\Omega_*$ is most
pronounced in the limit of weak repulsion, $G_T\gg2\sqrt{g}$.
Since the total resistance is the sum of $R_{T,\rm eff}$ and $R_D$,
cf.\ Eq.~(\ref{RA*}), the magnitude of the effect is determined
by the ratio $R_{T,\rm eff}/R_D$ which
has a maximum at $t\approx G_T/\sqrt{g}$.
An example of such a nonmonotonous
dependence of $G_A(\Omega_*)$ is shown in Fig.~\ref{F:ga_t_int}.
The curves differ by the ratio $t=R_D/R_T$ and correspond to
$G_T=10\sqrt{g}$ (i.~e., $\lambda_g/a=0.2$).

We emphasize that the described nonmonotonous behavior
has nothing to do with usual ``finite-bias"  conductance anomaly in N-S
structures with good interfaces~\cite{Art,StNaz}, which occur
at $(T, eV) \approx \ETh$ even in the absense of interaction in the N region.
In simple terms the origin of this new effect can be understood as
follows:
repulsion in the normal metal produces a superconductive
``gap function'' in the normal conductor, $\Delta_N$, with the negative
(compared to $\Delta_S$ in a superconductor) sign. Due to its opposite
sign, $\Delta_N $ decreases the conductance of the structure, therefore
any
decoherence that reduces $\Delta_N$ leads to the increase of the
conductance.

\section{Noise}

Now we turn to the analysis of the noise function $P_S$.
For a noninteracting system in the zero-energy limit
it is given by~\cite{DeJong,SLF}
\be
  P_S^{(0)}(t) = 1 +
  \frac
    {1+\Theta(t)\tan\Theta(t)+3\Theta(t)\cot\Theta(t)}
    {2[1+\Theta(t)\tan\Theta(t)]^4} .
\label{PS}
\ee
Eq.~(\ref{PS}) describes a crossover from the Poissonian ($P_S=3$)
to the sub-Poissonian ($P_S=1$)~\cite{1/3} character of the noise
as the system
evolves from the tunnel ($t\ll1$) to the diffusive ($t\gg1$) limits.

\begin{figure}
\refstepcounter{figure} \label{F:diagram}
\vspace*{1mm}
\epsfxsize=70mm
\centerline{\epsfbox{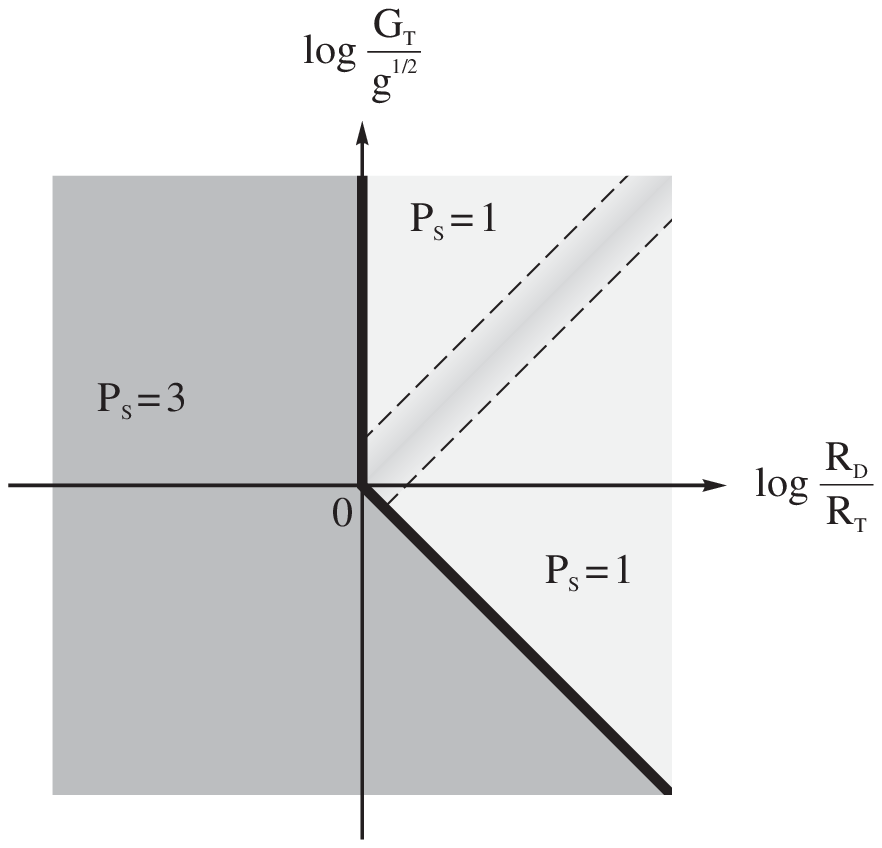}}
\vspace{1mm}
\small FIG.\ \arabic{figure}.
Schematic map of the noise coefficient $P_S$
as a function of the tunnel and diffusive resistances in the presence
of Cooper repulsion $\lambda=\lambda_g$. Dark area refers to
the tunnel limit $P_S\approx 3$, whereas light regions correspond to
the diffusive regime with $P_S \approx 1$.
\end{figure}

In the presence of interaction, $P_S$ can be estimated, qualitatively,
by comparing the effective tunneling resistance, $R_{T,\rm eff}$,
with the diffusive resistance, $R_D$.
For $R_{T,\rm eff} \gg R_D$, $P_S \approx 3$,
while in the opposite case $P_S \approx 1$.
If we again assume that $\lambda(\zeta)=\lambda_g$,
then $P_S$ will become a function of two parameters:
$t=R_D/R_T$ and $\lambda_g/a=2\sqrt{g}/G_T$.
The boundaries between the regions with $P_S=3$ and $P_S=1$
on the plane ($\log t$, $\log(G_T/\sqrt{g})$) are sketched
in Fig.~\ref{F:diagram}.

In the case of strong repulsion, $G_T\ll\sqrt{g}$,
$R_{T,\rm eff}$ very quickly (at $t\sim G_T/\sqrt{g}$)
saturates at $R_{T,\rm eff}(\infty) \equiv (4\sqrt{g}/G_T)R_T$.
Then one obtains~\cite{SLF}
$P_S^{\rm int} = P_N[R_D/R_{T,\rm eff}(\infty)]$,
where $P_N(t)=1+2/(1+t)^3$ is the noise function~\cite{Nazarov94T,SLF}
for the system with the N island.
Thus, in the limit $G_T\ll\sqrt{g}$,
the crossover between the tunnel and diffusive character
of noise is shifted to $t\sim \sqrt{g}/G_T \gg 1$,
see Fig.~\ref{F:diagram}.

\begin{figure}
\refstepcounter{figure} \label{F:lam_shot}
\vspace{1mm}
\epsfxsize=74mm
\centerline{\epsfbox{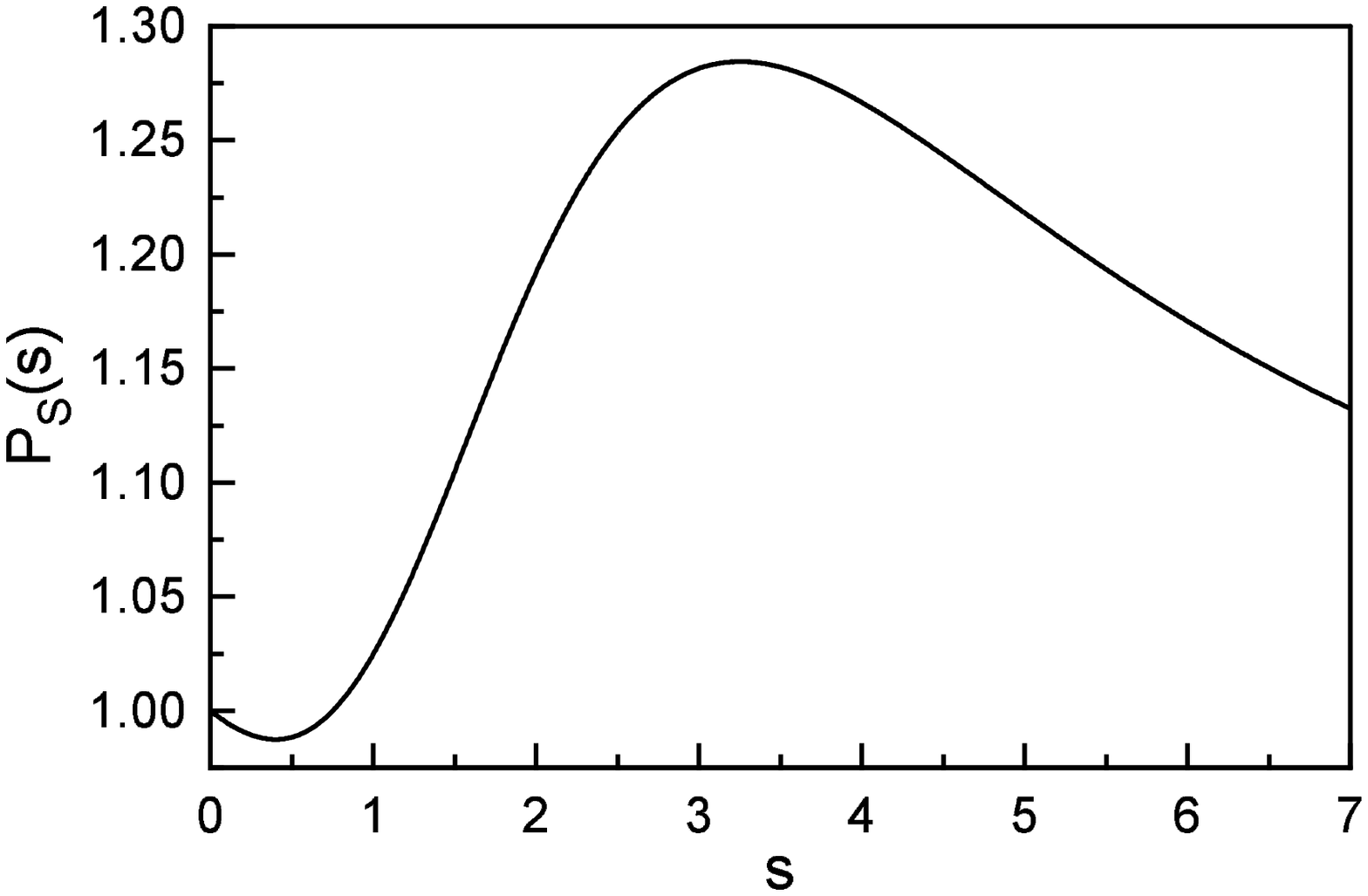}}
\small FIG.\ \arabic{figure}.
Noise function $P_S(s)$ vs.\ $s=2\sqrt{g}/G_D$
for the case of weak interaction, $G_T\gg\sqrt{g}$.
\end{figure}

In the limit of weak repulsion, $G_T\gg\sqrt{g}$, the situation is more
interesting. For $t\sim 1$, interaction corrections can be neglected
and $P_S$ is given by the noninteracting expression (\ref{PS}).
So, at $t\sim 1$, $P_S$ decreases from 3 to 1, the corresponding
boundary being shown in Fig.~\ref{F:diagram}.
Later, at $t\sim G_T/\sqrt{g}\gg1$
(when $R_D\sim\hbar/e^2\sqrt{g}$)
interaction corrections become relevant.
In this region, $R_{T,\rm eff}$ is of the order of $R_D$,
and one may anticipate that $P_S$ will deviate from 1.
For even larger $t$ when resistance is dominated by the diffusive
conductor,
$P_S$ will eventually reduce down to 1.
This crossover region
is marked in Fig.~\ref{F:diagram} by the dashed lines.
The function $P_S$ in the crossover region $t\sim G_T/\sqrt{g}\gg1$
obtained by numerical solution of Eq.~(\ref{urg})
is plotted in Fig.~\ref{F:lam_shot} as a function of
$s=\lambda_g\zeta=(\ln L/d)/\pi\sqrt{g} = 2(e^2/\hbar) R_D \sqrt g$.
It has a minimum $P_S=0.99$ at $s=0.40$
and a maximum $P_S=1.28$ at $s=3.25$.

Finally, we consider the $\Omega_*$ dependence of the noise power.
In the case of weak interaction, $P_S$ decreases (at $t,t_* \gg 1$)
with the increase of the Cooperon decoherence energy scale $\Omega_*$,
as if the system is becoming more diffusive.
This trend is opposite to what one has in the noninteracting case
when the increase of $\Omega_*$ drives the system toward the tunnel
limit, thus, increasing $P_S$.
In the limit of strong repulsion, $G_T\ll\sqrt{g}$, the zero-energy
noise function $P_S(t)$ exhibits a crossover from the tunnel to
diffusive
regimes at $t\sim \sqrt{g}/G_T\gg1$.
Nevertheless, upon increase of $\Omega_*$ the function $P_S(t,t_*)$
remains $t_*$ independent down to much smaller values of
$t_*\sim G_T/\sqrt{g}\ll1$ which correspond to the
energy scales $\ln(\omega_d/\Omega_*) \simeq 2\pi\sqrt{g}$.

\section{Conclusion}

We developed the Proximity Action functional method
able to describe quantum charge transport in mesoscopic superconductive
structures in the presence of interaction in the normal metal.
New method is applied to the study of
the effect of repulsion in the Cooper channel upon Andreev conductance
and noise in 2D N-S structures.  Interaction corrections scale
as $g^{-1/2}\ln(L/d)$ and lead to nonmonotonous dependence of both
the conductance and the Fano factor  upon temperature, voltage and
magnetic field.

\acknowledgements

We are grateful to G.~B.~Lesovik and Yu.~V.~Nazarov for useful
discussions. This research was supported by the NSF grant DMR-9812340
(A.~I.~L.), RFBR grant 98-02-16252, NWO-Russia collaboration grant,
Swiss NSF-Rus\-sia collaboration grant 7SUPJ062253.00,
and by the Russian Ministry of Science via the project
``Mesoscopic electron systems for quantum computing"
(M.~V.~F. and M.~A.~S.).


\end{multicols}

\end{document}